%% file: main.tex
\documentclass[
    prl,
    twocolumn,
    superscriptaddress,
    amsmath,
    amssymb,
    aps,
    floatfix
]{revtex4-2}
\usepackage{graphicx}
\usepackage{subcaption}
\usepackage{float}
\usepackage{bm}
\usepackage{xcolor}
\usepackage{hyperref}
\usepackage{comment}

\newcommand{\ua}{{\uparrow}}
\newcommand{\da}{{\downarrow}}
\newcommand{\vq}{{\vec{q}}}
\newcommand{\vk}{{\vec{k}}}
\newcommand{\vR}{{\vec{r}}}

\newcommand{\mysection}[1]{\noindent\textit{\textbf{#1}: }}

\begin{document}

\title{Flat-band formation and chiral superconductivity in driven topological insulators}

\author{Suman Jyoti De}
\affiliation{Physics Department, McGill University, Montr\'eal, QC, Canada H3A 2T8}
\author{Leo Goutte}
\affiliation{Institute of Physics and Center for Quantum Science and Engineering, Ecole Polytechnique F\'ed\'erale de Lausanne, CH-1015 Lausanne, Switzerland}
\author{Kartiek Agarwal}
\affiliation{Material Science Division, Argonne National Laboratory, Lemont, IL, USA 60548}
\author{T. Pereg-Barnea}
\affiliation{Physics Department, McGill University, Montr\'eal, QC, Canada H3A 2T8}

\date{\today}

\begin{abstract}
We demonstrate that circularly polarized light can be used to Floquet-engineer nearly flat or Mexican-hat like electronic bands on the surface of three-dimensional topological insulators (3D TIs), which under suitable conditions, can support topological superconductivity via purely repulsive Coulomb interactions. The driving acts not merely by gapping out the Dirac cone on the surface of the 3D TI, but can be used to diminish, and even flip in sign, the intrinsic curvature of the surface state dispersion away from the Dirac point. Using parameters for canonical 3D TIs, we find that the flat band limit is attained for reasonable electric fields and the bands realized by changing the strength of the driving field have a similar energetic and spatial profile to those obtained in rhombohedral graphene under varying displacement field, where the case for superconductivity with purely repulsive interactions has recently been made. We find that, with the aid of appropriately placed screening metallic gate, one can obtain $T_c \sim 7 $K in this setup while avoiding Wigner crystallization for low electron densities in the range of $10^{11}-10^{12}/\text{cm}^2$. 
\end{abstract}

\maketitle

\mysection{Introduction} 
Systems with nearly dispersionless electronic bands provide a fertile setting for strongly correlated quantum phases as electron-electron interactions dominate over the kinetic energy. Such flat bands can arise through a variety of mechanisms, ranging from single-particle interference in certain lattices to engineered electronic structures \cite{Artifical_flatband_Leykam2018, Flatband_ferromagnetism_Tasaki1998, Topological_Flatbands_Sarma2011, Flatbands_dicelattice_Wang2011, FCI_Shivaji2013}. For instance, in magic-angle twisted bilayer graphene and related moiré materials, the interplay between interlayer hybridization and larger moir\'e unit cell near the magic angle produces extremely narrow electronic bands, with superconducting phases appearing in proximity to correlated insulating states \cite{Pablo_MATBG_CI2018, Andrei_TBG2019, Cory_TBG2019, Efetov_MATBG2020, MacDonald_TBG2011, Fu_Chiral_SC2025}. More recently, these systems have also been shown to host fractional Chern insulating phases \cite{FCI_MoTe2_Shan2023, FCI_MoTe2_Lu2025}, highlighting the remarkable diversity of interaction-driven phenomena enabled by engineered flat bands.

\begin{figure}[h]
    \centering
    \begin{subfigure}{0.95\linewidth}
        \caption{}
        \includegraphics[width=\linewidth]{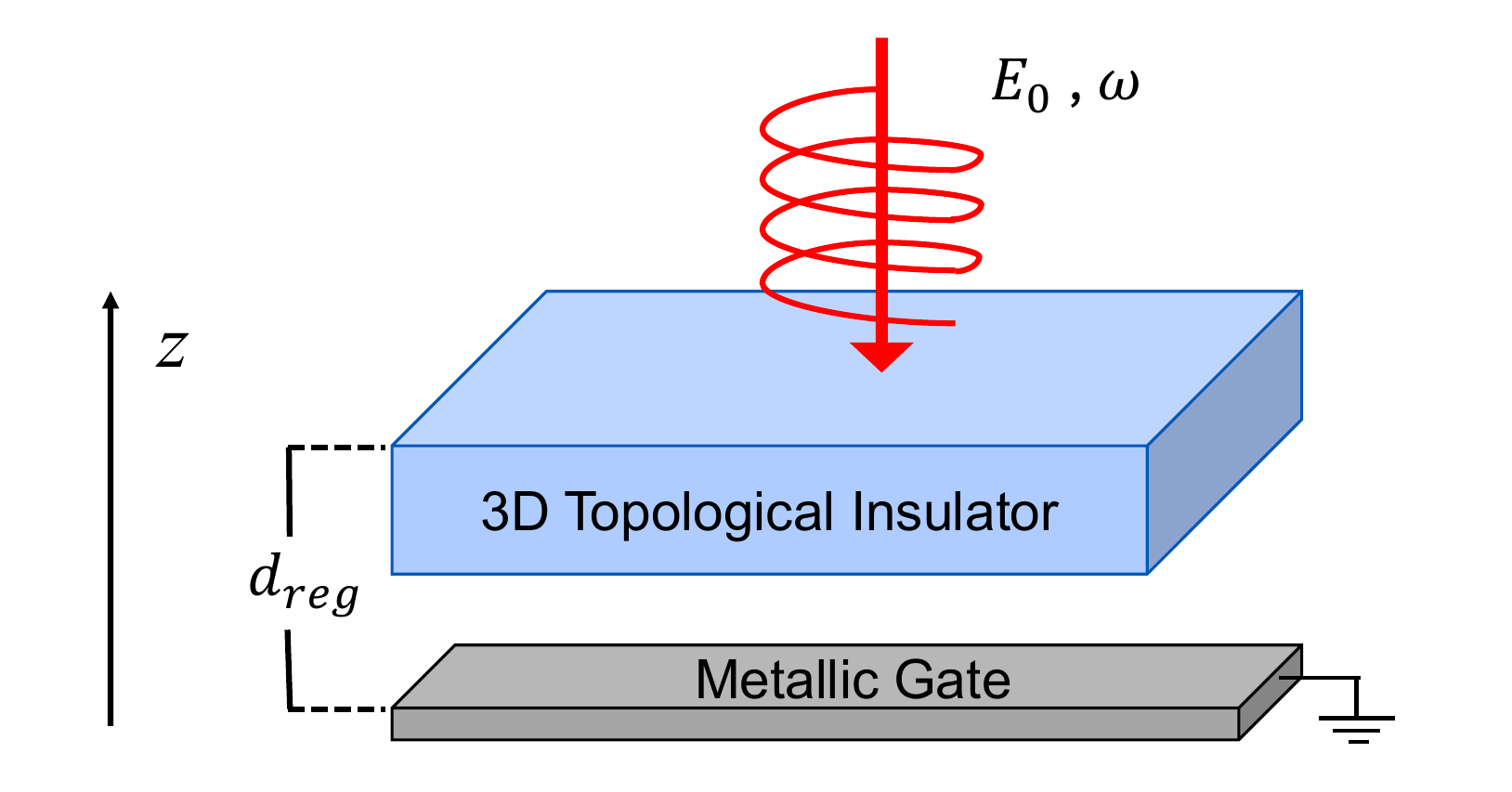}
    \end{subfigure}
    \begin{subfigure}{0.95\linewidth}
        \caption{}
        \includegraphics[width=\linewidth]{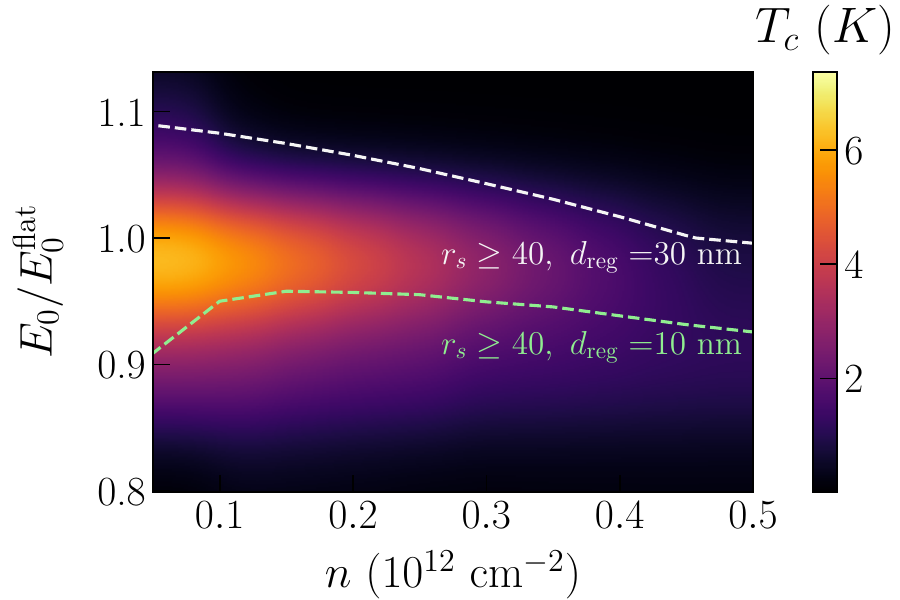}
    \end{subfigure}
    \caption{(a) Experimental set-up. (b) Phase diagram reflecting the superconducting critical temperature $T_c$ and gas parameter $r_s$ at different drive amplitudes $E_0$ and electron densities $n$ obtained by solving the gap equation for $l =1$, with $E_0^{\mathrm{flat}}=\frac{\hbar \omega}{e} \sqrt{\frac{\hbar \omega}{2|D|}}$. The region below dashed lines corresponds to $r_s \geq 40$ (suggestive of a Wigner-crystal phase) for the noted $d_\mathrm{reg}$.}
    \label{fig:set_up_and_Tc}
\end{figure}

A complementary route to band structure control is provided by Floquet theory, in which a system, driven periodically at high frequency, is mapped \cite{Shirley_Floquet1965, Effective_Floquet_Ha_Fishman2003, Effective_Floquet_Ha_Goldman2014, Floquet_quantum_matter_Goldman2015, Effective_Floquet_Ha_Eckardt_2015} onto an effective time-independent Hamiltonian with renormalized bands. In many cases, the resulting quasi-energy bands within a single Brillouin zone in frequency space differ significantly from their equilibrium counterparts, thus providing a purely dynamical handle on band engineering without altering material composition \cite{Lindner_Floquet_TI2011, Oka_Floquet2019, Rudner_Floquet_review2020, Zhou_light_induced_phenomena2022}. A paradigmatic realization is the irradiation of graphene with circularly polarized light \cite{Cavalleri_AQHEgraphene2020}, which gaps the Dirac cone via a photon-dressed effective next-nearest-neighbor hopping \cite{PhotovoltaicHE_graphene_Aoki2009, Eugene_AQHEgraphene2011, Irradiated_graphen_Balseiro2014, Floquet_Dirac_cone_Tami2016} and drives the system into a topological phase with a nonzero Chern number. 

In this letter, we propose using high-frequency periodic driving to generate narrow width electronic bands by irradiating circularly polarized light on the surface of three dimensional topological insulators (3D TIs) that engender a surface Dirac mode; see Fig.~\ref{fig:set_up_and_Tc}(a). Key to this proposal is the fact that, away from the Dirac point, higher-order momentum terms become relevant. As we show, driving not only gaps the Dirac cone into two bands as expected, but at the appropriate intensity, cancels these higher-order terms in an asymmetric fashion---exacerbating curvature in one, say upper band, and quenching it in the other, say lower band. The curvature of the lower band can then be reduced to zero, dispersing at only $\mathcal{O} (k^4)$, or even be made negative, realizing a Mexican hat-like shape. Moreover, the bands are approximately pseudospin polarized, potentially allowing for topological superconductivity.

Indeed, recent theoretical work on superconductivity in rhombohedral multilayer graphene has argued that such Mexican hat shaped bands can provide a favorable setting for topological superconductivity~\cite{Fu_Chiral_SC2025}. The enhanced density of states dramatically strengthens Thomas-Fermi screening, suppressing long-range repulsion of the Coulomb interaction while leaving intact its characteristic momentum dependence arising from the $2 k_F$ non-analyticity of the electronic susceptibility. The screened interaction can then become attractive at short distances, which combined with pseudospin polarization, leads to odd parity superconductivity. 

Motivated by these developments, we adapt the analysis of Ref.\cite{Fu_Chiral_SC2025} to the present setting of Floquet-engineered flat bands on the surface of 3D TIs. In the present setting, approximate spin polarization in the gapped Floquet bands is naturally achieved via strong spin orbit coupling. Metallic gates are further employed to control the strength and range of the interactions, and as we discuss, play a crucial role in promoting superconductivity over the insulating Wigner-crystal (WC) phase. We show that, for experimentally accessible drive frequencies and intensities, and with band structure parameters for canonical 3D TIs, topological superconductivity can be obtained for a range of electron densities.

\begin{figure*}[t]
    \centering
    \begin{subfigure}{0.33\linewidth}
        \caption{}
        \includegraphics[width=\linewidth,height=4cm]{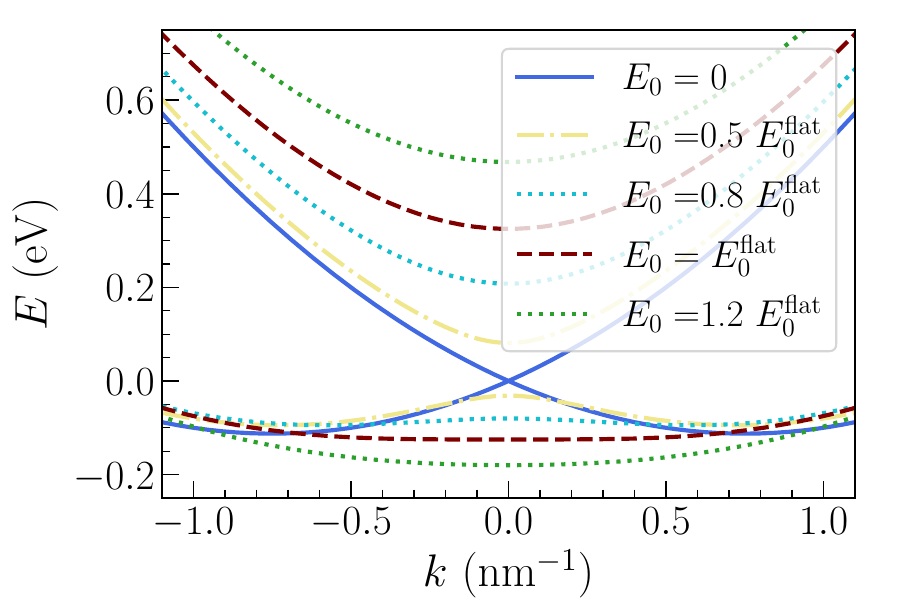}
    \end{subfigure}
    \hspace{-0.15cm}
    \begin{subfigure}{0.33\linewidth}
        \caption{}
        \includegraphics[width=\linewidth,height=4cm]{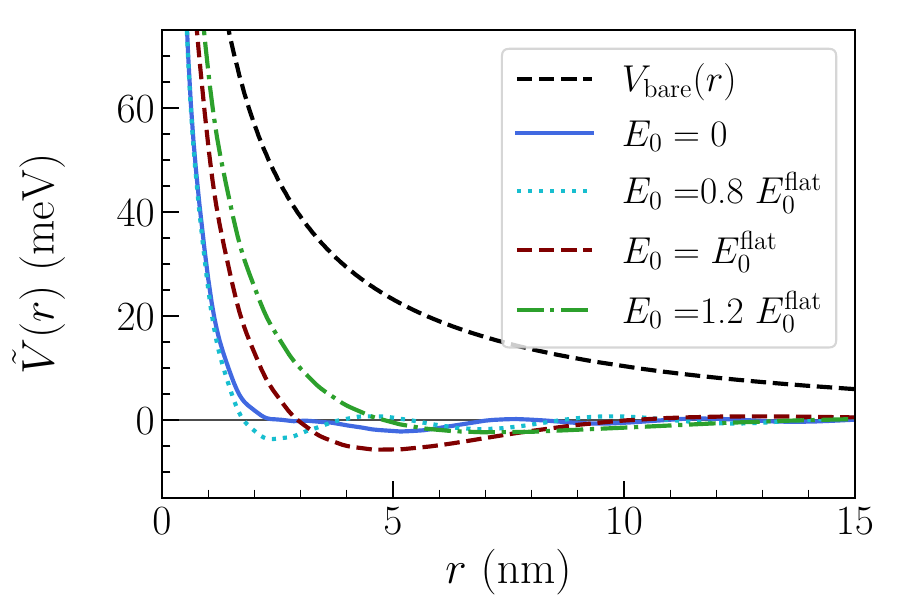}
    \end{subfigure}
    \hspace{-0.15cm}
    \begin{subfigure}{0.33\linewidth}
        \caption{}
        \includegraphics[width=\linewidth,height=4cm]{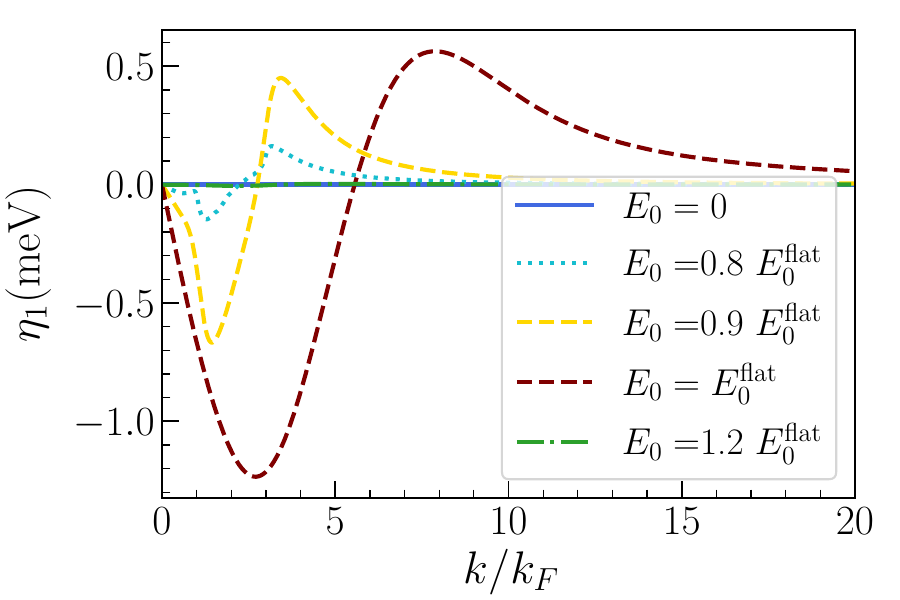}
    \end{subfigure}
    \caption{(a) Dispersion of the Floquet bands with increasing electric field strength $(E_0)$. For $E_0 = E_0^{\mathrm{flat}}$, the lowest Floquet band becomes nearly dispersionless near the Dirac point. (b) Real-space profile of screened Coulomb interaction and (c) radial momentum dependence of the self-consistent superconducting order parameter with orbital angular momentum $\ell=1$ at electron density $n=0.1 \times 10^{12} ~\mathrm{cm^{-2}}$ and $d_\mathrm{reg}=30$ nm. $V_\mathrm{bare}(r)$ is the unscreened Coulomb interaction.}
    \label{fig:Ek_Vr_Delta}
\end{figure*}

\mysection{Model} We consider an effective $2\times2$ low-energy model \cite{3DTI_Fu2007, 3DTI_Kane2007, 3DTI_Zhang2009, 3DTI_Zhang2010} describing the dispersion near the Dirac point on the surface of a 3D TI \cite{Bi2Se3_expt_2009, Bi2Te3_expt_2009, Hasan_Kane_TI2010}. Up to $\mathcal{O}(k^2)$, the system is described by the Hamiltonian 
\begin{align}
    \mathcal{H}_0= \sum_\vk c_\vk^\dagger \left[ \hbar v_F \left(\vk \cdot \vec{\sigma}\right)+D k^2 \sigma_0 \right] c_\vk ~,~c_\vk^\dagger=(c_{\vk \ua}^\dagger ~ c_{\vk \da}^\dagger) 
    \label{eq:continuum_H0_hamiltonian}
\end{align}
with $\vk=(k_x,k_y),~ k^2=k_x^2+k_y^2$ and $\sigma_i$, $i = 0, x, y, z$ are the Pauli matrices acting in the pseudospin space. For canonical 3D TIs such as Bi$_2$Se$_3$, Bi$_2$Te$_3$, Sb$_2$Te$_3$ that we consider, to a good approximation, the pseudospin $\vec{\sigma}$ aligns with the physical spin orientation. 

Under the influence of circularly polarized light, one obtains an effective time-independent Floquet Hamiltonian up to order $\frac{e A_0 v_F }{\hbar \omega}$ as 
\begin{align}
    \mathcal{H}_0^F=\sum_\vk c_\vk^\dagger \bigg[ \hbar v_F \left(\vk \cdot \vec{\sigma}\right) + D\left(k^2 + \frac{e^2 A_0^2}{\hbar^2} \right)\sigma_0  \nonumber \\
    - \frac{v_F^2 e^2 A_0^2}{\hbar \omega} \sigma_z \bigg] c_\vk 
    \label{eq:continuum_floquet_hamiltonian}
\end{align}

where $A_0=\frac{E_0}{\omega}$, $E_0, \omega$ are the driving field amplitude and frequency, respectively, and $e$ is the electron charge. See supplementary material (SM) for the derivation of the Floquet Hamiltonian as well as justification for retaining terms up to the first non-trivial order in the inverse frequency expansion. Furthermore, for a driven topological insulator, Refs.~\cite{Dissipative_Floquet_system_Karthik_1, Dissipative_Floquet_system_Karthik_2} argue that sufficiently strong coupling to phonons generally leads to a steady state with the electronic population closely given by the static thermal distribution. In our set-up, we will assume such a steady state is achieved, and therefore restrict our analysis to a purely static system interacting via Coulomb interaction.

In what follows, we will be interested in situations where the electron density is small, and electrons merely populate the lower band. We can thus limit ourselves to an analysis of interaction effects in the lower band, which has the dispersion
\begin{align}
    \varepsilon_{\vk} = D\left(k^2 + \frac{e^2 A_0^2}{\hbar^2} \right) - \sqrt{\hbar^2 v_F^2 k^2+ \left(\frac{v_F^2 e^2 A_0^2}{\hbar \omega}\right)^2}
    \label{eq:lowerband_dispersion}
\end{align}
Assuming $D > 0$, for $E_0$ close to $E_0^{\mathrm{flat}}=\frac{\hbar \omega}{e} \sqrt{\frac{\hbar \omega}{2|D|}}$, the bandwidth of this lower band reduces significantly, as shown in Fig.\ref{fig:Ek_Vr_Delta}(a). Precisely at $E_0 = E_0^{\mathrm{flat}}$, the band disperses only at $\mathcal{O} \left(k^4 \right)$; we will refer to this as the flat band limit. For $E_0 < E_0^{\mathrm{flat}}$, the Fermi surface assumes the shape of an annulus; thus here the drive strength plays a similar role to the displacement field in Ref.~\cite{Fu_Chiral_SC2025}, inducing band flattening and a Lifshitz transition.  

For typical 3D TIs such as Bi$_2$Se$_3$, Bi$_2$Te$_3$, Sb$_2$Te$_3$; the parameter values are $v_F \approx 5 \times 10^5~ \mathrm{m/s}$ and $|D| \approx 0.1-0.3~ \mathrm{eV} (\mathrm{nm^2})$. Among these, Bi$_2$Se$_3$ has the lowest dielectric constant \cite{Bi_dielectric_prop_Furdyna2019}, which favours a higher superconducting critical temperature; thus, in this work we consider parameter values specific to Bi$_2$Se$_3$. 
We note that using the parameters for Bi$_2$Se$_3$ \cite{3DTI_Zhang2009, 3DTI_Zhang2010, Bi2Se3_expt_2009}, $\hbar v_F = 0.3~ \mathrm{eV} (\mathrm{nm})$, $D = 0.2~ \mathrm{eV} (\mathrm{nm^2})$, and drive frequency $\omega = 2\pi \cdot 50 $ THz, as has been used to investigate Floquet physics in this material \cite{Rudner_Floquet_review2020, Zhou_light_induced_phenomena2022}, one obtains $E_0^{\mathrm{flat}} \simeq 1.4 \times 10^8 ~\mathrm{V/m}$ which is within experimental reach \cite{Wang_expt_Floquet_TI2013, Zhou_light_induced_phenomena2022}.

\mysection{Screened Coulomb interaction} The full interaction between electrons in the lower Floquet band takes the general form
\begin{align}
    \mathcal{H}_{int}= \frac{1}{2A} \sum_{\vk_1,\vk_2,\vq} V(\vq)~ \langle u_{\vk_1} | u_{\vk_1-\vq} \rangle \langle u_{\vk_2} | u_{\vk_2+\vq} \rangle \nonumber \\ \times ~\psi_{\vk_1}^\dagger \psi_{\vk_2}^\dagger \psi_{\vk_2+\vq} \psi_{\vk_1-\vq}
    \label{eq:full_int_hamiltonian}
\end{align}
where $\psi_{\vk}^{\dagger}$ creates an electron in the lower Floquet state $|u_{\vk}\rangle$. We find that finite superconducting order prevails close to the flat-band limit; in this limit, we can approximate the single particle wavefunction for $A_0=A_0^\mathrm{flat}=\frac{\hbar}{e} \sqrt{\frac{\hbar \omega}{2|D|}}$, up to $\mathcal{O}(k^2)$ as
\begin{align}
|u_{\vk}\rangle \approx \frac{1}{\sqrt{1+\frac{\hbar^2 k^2}{4m^2_\mathrm{flat} v^2_F}}}
 \begin{pmatrix}
    1  \\
    \frac{-\hbar v_Fk}{2 m_\mathrm{flat} v^2_F}~ e^{-i \theta_k}
\end{pmatrix}
\label{eq:single_particle_wf}
\end{align}
with $m_\mathrm{flat}=\frac{\hbar^2}{2 |D|}$ and $\vk=(k\cos{\theta_k}, k\sin{\theta_k})$.
For electron densities considered here, close to the flat-band limit, electrons populate states near the dispersionless region with largest occupied Fermi momentum $k_F \approx 0.3~ \mathrm{nm^{-1}}$, so using form of the single-particle wavefunction in Eq.(\ref{eq:single_particle_wf}), we can assume form factors $\langle u_{\vk+\vq} | u_\vk \rangle = 1 $, with neglecting corrections of order $\frac{\hbar^2 k_F^2}{4 m^2_\mathrm{flat} v_F^2 } \ll 1$ for $q \simeq k_F$. Thus, we can ignore the projection factors in Eq.(\ref{eq:full_int_hamiltonian}) and primarily work with the bare interaction
\begin{align}
    \mathcal{H}_{int}= \frac{1}{2A} \sum_{\vk_1,\vk_2,\vq} V(\vq) ~\psi_{\vk_1}^\dagger \psi_{\vk_2}^\dagger \psi_{\vk_2+\vq} \psi_{\vk_1-\vq}
    \label{eq:continnum_int_hamiltonian}
\end{align}
Hereon, we present numerical results with the interactions given by the Hamiltonian in Eq.(\ref{eq:continnum_int_hamiltonian}). See SM for more discussion on the effect of retaining projection terms.

We model the Coulomb interaction between electrons by the Rytova-Keldysh potential \cite{Rytova2020, Keldysh1979}
\begin{align}
    V(\vq) = \frac{e^2 \tanh[q d_{reg}]}{2 \epsilon \left(q + r_K q^2\right)}
    \label{eq:bare_Coulomb_interaction}
\end{align}
with $q$ being the magnitude of $\vq$, which is appropriate for electrons in a 2D surface state in proximity to metallic gates having a lower dielectric permittivity than the parent thin film material. $d_{reg}$ is distance of the metallic gate from the top surface of the 3D TI, $~\epsilon=\left(\epsilon_{\mathrm{TI}}+\epsilon_0\right)/2$ is the effective dielectric permittivity for the surface states in contact with vacuum, and $r_K$ is the Rytova-Keldysh parameter. For Bi$_2$Se$_3$, we use $\epsilon_{\mathrm{TI}}=20 \epsilon_0$ \cite{Bi_dielectric_prop_Furdyna2019, Bi2Se3_dielectric_prop_Shiyun2020} (with $\epsilon_0$ is the vacuum permittivity), and $r_K=1$ nm.

Near the center of the band in the flat-band limit, the pseudospin is polarized uniformly with corrections to order $\mathcal{O} (k^2/k_0^2)$, $k_0=\frac{\hbar v_F}{\sqrt{2}|D|}$; thus, spin is effectively frozen at temperatures $k_BT \ll D k_0^2$ ($\approx 0.1 ~\mathrm{eV}$, using the parameters for Bi$_2$Se$_3$). We thus ignore spin fluctuations and consider pairing as arising from charge fluctuations akin to the Kohn-Luttinger mechanism \cite{Kohn_Luttinger1965, KL_SC_Chubukov1993, KL_SC_Saurabh2013, Alexandrov2003book}, using screened Coulomb interaction. Using random phase approximation (RPA) \cite{Mahan1990book}, the effective screened interaction can be written as 
\begin{align}
   \tilde{V}(\vq) \overset{\mathrm{RPA}}{\approx} \frac{V(\vq)}{1 - V(\vq) \chi_c^0(\vq,0)}
    \label{eq:screened_Coulomb_interaction}
\end{align}
where $V(\vq)$, as shown in Eq.(\ref{eq:bare_Coulomb_interaction}), is the unscreened interaction, and the bare charge susceptibility is given by
\begin{align}
    \chi_c^0(\vq, i \Omega_n) = \frac{1}{A} \sum_{\vk} \frac{n_F(\varepsilon_{\vk}) - n_F(\varepsilon_{\vk+\vq})}{\varepsilon_{\vk} - \varepsilon_{\vk+\vq} + i \hbar \Omega_n}
    \label{eq:bare_chi}
\end{align}
$n_F(\varepsilon)$ is the zero-temperature electron population, given by the Fermi-Dirac distribution function.

The effective screened interaction profile in real space is found by Fourier transform and is shown in Fig.\ref{fig:Ek_Vr_Delta}(b). We see that screening strongly reduces Coulomb repulsion between electrons. Moreover, near $E_0 \approx E_0^{\mathrm{flat}}$ because of large density of states near the Fermi surface, we find that the screened interaction becomes significantly negative over a wide range of distances, thereby enabling superconductivity.

\mysection{Chiral superconductivity} We solve the mean-field self-consistency equation 
\begin{align}
    \Delta(\vk) = - \frac{1}{A} \sum_{\vk'} \tilde{V}(\vk-\vk') \langle \psi_{\vk'} \psi_{-\vk'} \rangle ~, \nonumber \\
    \langle \psi_{\vk} \psi_{-\vk} \rangle = \frac{\Delta(\vk)}{2 E_{\vk}} \tanh \left( \frac{E_{\vk}}{2k_B T} \right)
    \label{eq:MF_SC_eq}
\end{align}
with the quasi-particle energies
\begin{align}
    E_{\vk}=\sqrt{(\varepsilon_\vk - \mu)^2 + |\Delta(\vk)|^2}
    \label{eq:BDG_en}
\end{align}
to compute the superconducting (SC) order parameter $\Delta(\vk)$, and the corresponding critical temperature $T_c$.

Due to rotational symmetry in the problem, we decompose the effective interaction and order parameter into angular harmonics:
\begin{align}
    \tilde{V}_\ell(k, k') = \int_0^{2\pi} \frac{d\theta}{2\pi}~ e^{i\ell \theta}~ \tilde{V}(\vk-\vk') ~,~ \cos{\theta}=\frac{\vk.\vk'}{k~ k'} \nonumber \\
    \Delta(\vk) = \sum_\ell \eta_\ell(k)~ e^{i\ell \phi} ~,~ \tan{\phi} = \frac{k_y}{k_x}
    \label{eq:angular_decomposition}
\end{align}
where $k$ is the magnitude of momentum vector $\vk=(k_x, k_y)$. This allows us to solve for each angular momentum channel $(\ell)$ separately. See SM for the detailed calculations, where we show how we can obtain $T_c$ for each angular momentum channel.

As mentioned earlier, because of spin polarization of the lower Floquet band, only odd angular momentum pairings survive. Within our parameter regime, we find that the dominant angular momentum pairing is $\ell = \pm 1$ associated with chiral $p_x \pm i p_y$ SC order. SC order parameter for the angular momentum channel $\ell=1$ at zero temperature is then found by solving equation
\begin{align}
    \eta_\ell(k) = - \int_0^{\infty} \frac{k' dk'}{4 \pi} \frac{\tilde{V}_\ell(k, k') \eta_\ell(k')}{\sqrt{\left( \varepsilon_{k'}-\mu \right)^2 + \eta^2_\ell(k')}}
    \label{eq:self_eta_zero}
\end{align}
self-consistently and is shown in Fig.\ref{fig:Ek_Vr_Delta}(c). We get Eq.(\ref{eq:self_eta_zero}) straightforwardly from equations (\ref{eq:MF_SC_eq}, \ref{eq:BDG_en}) by using definitions in Eq.(\ref{eq:angular_decomposition}). The results for $\eta_\ell$ reveal strong momentum dependence and thus suggest the necessity of solving the gap equation over a wide range of momenta. We find that the electric field strength needs to be close to, and slightly lower than  the flat band limit, that is, $E_0 \lesssim E^{\text{flat}}_0$, to produce the strongest SC order; see Fig.~\ref{fig:set_up_and_Tc} also. This suggests a strong correlation between induced pairing and flat-band formation, and proximity to the Lifshitz transition to a Mexican hat like band.

As mentioned above, we use the bare Coulomb interaction (Eq.\ref{eq:continnum_int_hamiltonian}) in our self-consistent calculation of the superconducting order parameter and critical temperature.  Without any band information, this interaction does not favor a particular chirality. However, starting from the projected interaction, while keeping all the form factors as in Hamiltonian ~\ref{eq:full_int_hamiltonian}, which inherits the band chirality, we find that, for our parameter regime, the $l=-1$ angular momentum channel corresponding to $p_x - ip_y$ pairing is energetically favored and yields higher $T_c$; see SM for more details.

\mysection{Gas parameter and Experimental Implications} Assuming a large enough frequency $\left( \omega  \text{ s.t. }  \hbar \omega \approx 0.2 \mathrm{(eV)} \right)$ and with a tunable electric field strength $(E_0)$ of the driving field, we find the possibility of superconductivity (see Fig.\ref{fig:set_up_and_Tc}(b)) for a wide range of surface electron densities $n = 10^{11} - 10^{12} / \text{cm}^2$. As superconductivity emerges at relatively low densities, we examine competition with the WC phase by computing the gas parameter \cite{Giuliani2005book} $r_s = \frac{2 V_\mathrm{bare}(r_0)}{E_F(n)-E_\mathrm{bottom}}$. $V_\mathrm{bare}(|\vR|)$ is Fourier transformation of the bare electron interaction $V(|\vq|)$ shown in Eq.(\ref{eq:bare_Coulomb_interaction}) and $r_0=\sqrt{\frac{2}{\sqrt{3} n}}$ is typical distance between electrons in the WC phase. $E_F(n)$ is the Fermi-energy and $E_\mathrm{bottom}$ is band-bottom energy. As shown in Fig.\ref{fig:set_up_and_Tc}(b), we find that for $d_\mathrm{reg} \geq 30$ nm, the region with finite SC critical temperature $T_c$ is covered mostly by the region having $r_s \geq 40$, suggesting a strong possibility of realizing the WC phase \cite{Wigner1934, WC_Ceperley1978, WC_Ceperley1989, WC_Drummond2009}. Whereas by reducing $d_\mathrm{reg} \approx 10$ nm, we find that the region having $r_s \geq 40$ shrunk below the finite $T_c$ region. Thus, having a gate separation of $d_\mathrm{reg} \approx 10$ nm significantly enhances the possibility of realizing superconductivity in our system. On the contrary, we find that SC $T_c$ is weakly affected by $d_\mathrm{reg}$ and starts to decrease significantly only when $d_\mathrm{reg} \leq 5$ nm (see SM for a detailed comparison).

For a thin slab of 3D TI, when the slab thickness becomes comparable to the penetration depth of the surface states, wavefunctions localized on the top and bottom surfaces hybridize, opening a mass gap at the Dirac point. Spectroscopic measurements on Bi$_2$Se$_3$ have shown that this hybridization becomes significant for slab thicknesses $L \lesssim 6$nm \cite{Thin_slab_TI_Yi2010, Thin_slab_TI_Yusuke2010, Thin_slab_TI_Lu2010}. Thus, if we consider a metallic gate placed at a distance $d_\mathrm{reg} \approx 10$ nm from the top surface, provided the slab thickness satisfies $6\mathrm{nm} < L < d_\mathrm{reg}$, overlap between the top and bottom surface states remains negligible. This allows us to treat the two surfaces effectively decoupled and focus solely on the single Dirac cone of the top surface.

\mysection{Discussion} In this work, we show that circularly polarized light can be used to dynamically engineer nearly flat electronic bands on the surface of a three-dimensional topological insulator. By tuning the driving field amplitude to compensate for the band curvature near the Dirac point, the electronic bandwidth becomes strongly suppressed, thereby enhancing electron-electron interactions and enabling the emergence of a chiral superconducting state. Although originally introduced to describe the surface states of 3D TI, the continuum Hamiltonian studied here also captures, exactly or approximately, the low-energy physics of several other Dirac materials, including topological crystalline insulators \cite{Fu_TCI2011, Ando_TCI2015}, spin-orbit-coupled transition-metal dichalcogenides \cite{Xiao_TMD2012, Xu_TMDReview2014}, and graphene-based heterostructures with proximity-induced spin-orbit interactions \cite{Gmitra2015, Zhang2014}. Thus, our findings establish a general framework for exploring Floquet-engineered flat bands and interaction-driven phases across a wide range of materials.

\mysection{Acknowledgements} SJD and TPB benefit from their RQMP membership https://doi.org/10.69777/309032. We acknowledge the support of the Natural Sciences and Engineering Research Council of Canada (NSERC), NSERC CREATE/ 575280-2023 - Training in Materials for Quantum Technologies (MaQTech). SJD would also like to acknowledge the Digital Research Alliance of Canada for the use of clusters in numerical simulations. KA acknowledges funding from US Department of Energy, Office of Science, Basic Energy Sciences. 

\bibliographystyle{apsrev4-2}
\bibliography{floquet}

\clearpage
\onecolumngrid
\input{SM}

\end{document}

%% file: SM.tex
\begin{center}
{\large\bfseries Supplementary Material for: Flat-band formation and chiral superconductivity in driven topological insulators}
\end{center}


\section{Floquet Hamiltonian and flat-band condition}

In the presence of circularly polarized light $\vec{A}(t)=A_0(\cos\omega t,\sin\omega t,0)$, $\omega=\frac{2\pi}{T_p}$, $A_0=\frac{E_0}{\omega}$, with $E_0$ being the electric field strength, the effect can be taken into consideration by the Peierl's substitution in the Hamiltonian of Eq.(1) in the main text, where the momentum vector $\vk$ is replaced by the mechanical momentum $\vec{\Pi}(t)=\vk - \frac{e \vec{A}(t)}{\hbar}$.
The time-dependent Hamiltonian is therefore
\begin{align}
    \mathcal{H}_0(t)= \sum_\vk c_\vk^\dagger \left[\hbar v_F \big(\vec{\Pi}(t)\cdot \vec{\sigma}\big)+D~ \vec{\Pi}(t) \cdot \vec{\Pi}(t) \sigma_0 \right] c_\vk 
    \label{eq:H0_t}
\end{align}
We find the effective Floquet Hamiltonian using a well-studied inverse-frequency expansion \cite{Effective_Floquet_Ha_Goldman2014, Floquet_quantum_matter_Goldman2015, Effective_Floquet_Ha_Eckardt_2015, Universal_high-frequency_behaviorBukov2015, BW_expansion_Aoki2016},

\begin{align}
    \mathcal{H}^F_0=\sum_{\nu=0}^\infty H_F^{(\nu)}~, ~
    H_F^{(0)}=\hat{H_0}~, ~
    H_F^{(1)}=\sum_{n>0} \frac{\left[ \hat{H}_{n},\hat{H}_{-n} \right]}{n\hbar \omega}~, \nonumber \\
    H_F^{(2)}=\frac{1}{\hbar^2 \omega^2} \sum_{n \neq 0} \left( \frac{\left[ \hat{H}_{-n}, \left[\hat{H}_{0},\hat{H}_{n}\right] \right]}{2 n^2} + \sum_{n' \neq 0,n} \frac{\left[ \hat{H}_{-n'}, \left[\hat{H}_{n'-n},\hat{H}_{n}\right] \right]}{3 n n'} \right) ~, ...
    \label{eq:inverse_frequency_expansion}
\end{align}
where 
\begin{align}
    \hat{H}_n=\frac{1}{T} \int_0^T dt~ e^{-in\omega t} ~\mathcal{H}_0(t) 
    \label{eq:Hn_omega}
\end{align}
With this, up to first non-trivial order in the expansion parameter $\frac{e A_0 v_F }{\hbar \omega}$, we get an effective Floquet Hamiltonian as 
\begin{align}
    \mathcal{H}_0^F=\sum_\vk c_\vk^\dagger \bigg[ \hbar v_F \left(\vk \cdot \vec{\sigma}\right) + D\left(k^2 + \frac{e^2 A_0^2}{\hbar^2} \right)\sigma_0 
    -\frac{v_F^2 e^2 A_0^2}{\hbar \omega} \sigma_z \bigg] c_\vk 
    \label{eq:floquet_hamiltonian_1storder}
\end{align}

The eigenvalues of the Floquet Hamiltonian are
\begin{align}
    E_{\pm}=D\left(k^2 + \frac{e^2 A_0^2}{\hbar^2} \right) \pm \sqrt{\hbar^2 v_F^2 k^2+ \left(\frac{v_F^2 e^2 A_0^2}{\hbar \omega}\right)^2}
\end{align}
Let $\frac{e^2 A_0^2}{\hbar \omega}=m_0$, expanding for momentum $k \approx 0$ we get,
\begin{align}
   E_{\pm} \simeq  D\left(k^2 + \frac{e^2 A_0^2}{\hbar^2} \right) \pm m_0 v_F^2 \left( 1+\frac{\hbar^2 k^2}{2m_0^2 v_F^2} \right) + \mathcal{O} (k^4) \nonumber \\
   \simeq \frac{D e^2 A_0^2}{\hbar^2} \pm m_0 v_F^2 + k^2 \left( D \pm \frac{\hbar^2}{2m_0} \right)  + \mathcal{O} (k^4)
\end{align}
which reveals the flat band condition near $k \approx 0$ as $|D|=\frac{\hbar^2}{2m_0}$. To achieve the flat-band limit, the amplitude of the driving field needs to be tuned close to $\frac{e A_0}{\hbar}=\sqrt{\frac{\hbar \omega}{2|D|}}$.

\begin{figure}
    \centering
    \includegraphics[width=0.9\linewidth]{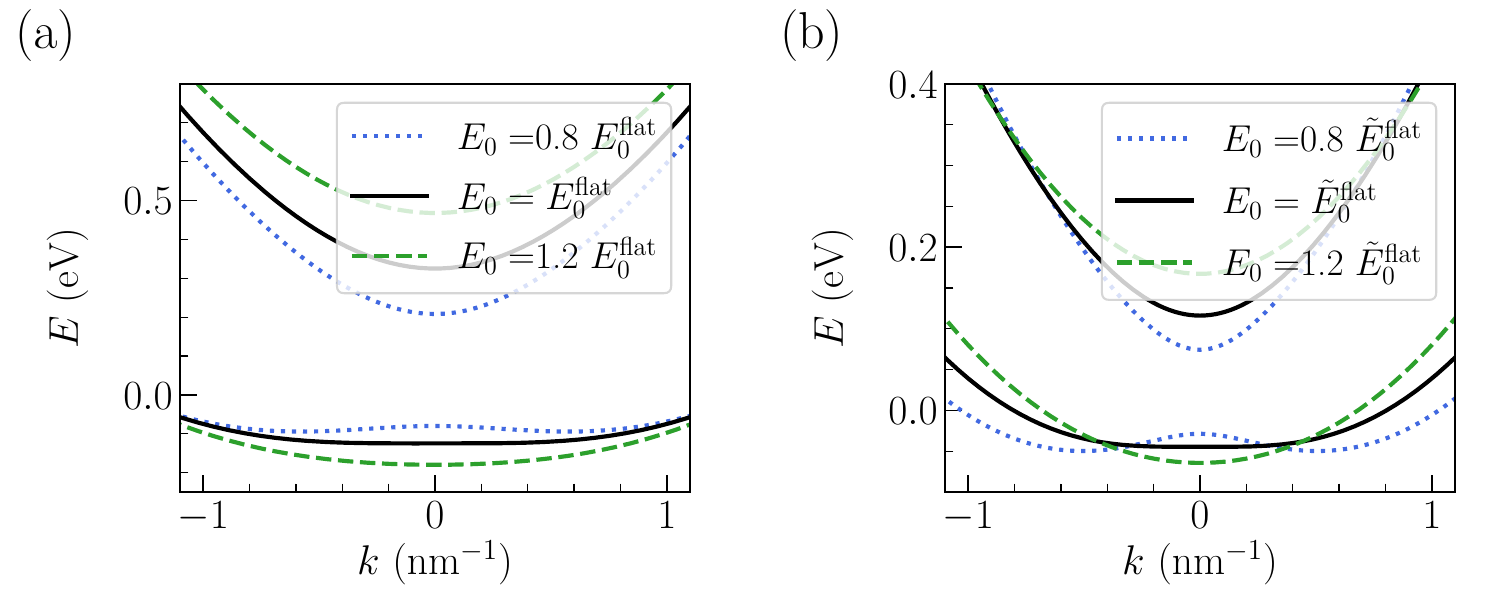}
    \caption{Comparison between bands close to flat-band limits $E_0^{\mathrm{flat}}$, $\tilde{E}_0^{\mathrm{flat}}$; with retaining terms up to (a) first order and (b) second order in the Floquet Hamiltonian with expansion \ref{eq:inverse_frequency_expansion}. See text for the definition of $E_0^{\mathrm{flat}}$ and $\tilde{E}_0^{\mathrm{flat}}$.}
    \label{fig:floquet_band_comparison}
\end{figure}

For the parameter values of Bi$_2$Se$_3$; $\hbar v_F = 0.3~ \mathrm{eV} (\mathrm{nm})$, $D = 0.2~ \mathrm{eV} (\mathrm{nm^2})$, and with drive frequency $\omega = 2\pi \cdot 50 $ THz, we get an expansion parameter value $\frac{e A_0 v_F }{\hbar \omega} \approx 1$, close to the flat-band limit $\frac{e A_0}{\hbar}=\sqrt{\frac{\hbar \omega}{2|D|}}$. Thus, close to this limit it is not straightforward to neglect higher-order terms.

Using equations \ref{eq:H0_t} and \ref{eq:Hn_omega} , in the inverse-frequency expansion \ref{eq:inverse_frequency_expansion} we get the next order term as
\begin{align}
   H_F^{(2)} = - \sum_\vk \frac{e^2 A_0^2}{\hbar^2} \times \frac{\hbar^3 v_F^3}{\hbar^2 \omega^2}~ c_\vk^\dagger \left( \vk.\vec{\sigma} \right) c_\vk
\end{align}

If we include this term in the effective Floquet Hamiltonian \ref{eq:floquet_hamiltonian_1storder}, the resultant Hamiltonian will be
\begin{align}
    \mathcal{H}_0^F = \sum_\vk c_\vk^\dagger \bigg[ \hbar \tilde{v}_F \left(\vk.\vec{\sigma}\right) + D\left(k^2 + \frac{e^2 A_0^2}{\hbar^2} \right)\sigma_0 
    -\frac{v_F^2 e^2 A_0^2}{\hbar \omega} \sigma_z \bigg] c_\vk 
    \label{eq:floquet_hamiltonian_2ndorder}
\end{align}
with 
$\tilde{v}_F = v_F \left(1 - \frac{e^2 A_0^2 v_F^2}{\hbar^2 \omega^2} \right)$. For this Hamiltonian, we get the revised flat band condition near $k \simeq 0$ as $|D|=\frac{\hbar^2 \tilde{v}_F^2}{2m_0 v_F^2}$, and to achieve this amplitude of the driving field needs to be tuned close to $\frac{eA_0}{\hbar}=\frac{\sqrt{\frac{2|D|}{\hbar \omega}+\frac{4 \hbar^2 v^2_F}{\hbar^2 \omega^2}}-\sqrt{\frac{2|D|}{\hbar \omega}}}{\frac{2 \hbar^2 v^2_F}{\hbar^2 \omega^2}}$.

In figure \ref{fig:floquet_band_comparison}, we show a comparison between the effective Floquet bands when we tune amplitude of the driving field $E_0$ close to flat-band limits $E_0^{\mathrm{flat}}=\frac{\hbar \omega}{e} \sqrt{\frac{\hbar \omega}{2|D|}}$, $~\tilde{E}_0^{\mathrm{flat}}=\frac{\hbar \omega}{e} \left( \frac{\sqrt{\frac{2|D|}{\hbar \omega}+\frac{4 \hbar^2 v^2_F}{\hbar^2 \omega^2}}-\sqrt{\frac{2|D|}{\hbar \omega}}}{\frac{2 \hbar^2 v^2_F}{\hbar^2 \omega^2}} \right)$ by retaining terms up to first and second order in the expansion of the Floquet Hamiltonian $\mathcal{H}_0^F$ (\ref{eq:inverse_frequency_expansion}), respectively. This clearly shows that even with the second-order term, the regions in the Brillouin zone having a flat band still survive. 

To illustrate the possibility of forming a flat band using periodic driving, it is thus sufficient to consider the effective Floquet Hamiltonian in Eq.(\ref{eq:floquet_hamiltonian_1storder}) and restrict terms up to the first non-trivial order in the expansion parameter. Materials having a smaller value of $v_F$, a larger value of $D$ and with a larger driving frequency $\omega$ (such that the amplitude of the driving field around the flat-band limit is smaller than the breakdown voltage of the material), will make the expansion parameter $\frac{e A_0 v_F }{\hbar \omega} = \frac{\hbar v_F}{\sqrt{2|D|~ \hbar \omega}} < 1$, close to the flat-band limit $\frac{e A_0}{\hbar}=\sqrt{\frac{\hbar \omega}{2|D|}}$. For those materials, it will be straightforward to neglect the higher-order terms beyond the first non-trivial order.

\section{Mean-field equation for the superconducting order}

To determine the superconducting (SC) order parameter, we solve the mean-field equation \cite{Alexandrov2003book} :
\begin{align}
    \Delta(\vk) = - \frac{1}{A} \sum_{\vk'} \tilde{V}(\vk-\vk') \frac{\Delta(\vk')}{2 E_{\vk'}} \tanh \left( \frac{E_{\vk'}}{2k_B T} \right)
    \label{eq:MF_SC_eq}
\end{align}
with the quasi-particle energies
\begin{align}
    E_{\vk}=\sqrt{(\varepsilon_\vk - \mu)^2 + |\Delta(\vk)|^2}
    \label{eq:BDG_en}
\end{align}
$\varepsilon_{\vk}$ is the single-particle energy, and $\tilde{V}(\vq)$ is the effective screened interaction found using the RPA approximation.

We follow a similar numerical method as explained in Ref.~\cite{Fu_Chiral_SC2025} while solving the mean-field equation. Due to rotational symmetry in the problem, we can decompose the effective interaction and order parameter into angular harmonics
\begin{align}
    \tilde{V}_\ell(k, k') = \int_0^{2\pi} \frac{d\theta}{2\pi}~ e^{i\ell \theta}~ \tilde{V}(\vk-\vk') ~,~ \cos{\theta}=\frac{\vk.\vk'}{k~ k'} \nonumber \\
    \Delta(\vk) = \sum_\ell \eta_\ell(k)~ e^{i\ell \phi} ~,~ \tan{\phi} = \frac{k_y}{k_x}
    \label{eq:angular_decomposition}
\end{align}
where $k$ is the magnitude of momentum vector $\vk=(k_x, k_y)$. With these and after some straightforward algebra, the angular momentum channels decouple, and  we get
\begin{align}
    \eta_\ell(k) = - \int_0^\infty \frac{k' dk'}{4 \pi} \tilde{V_\ell}(k, k') \frac{\eta_\ell(k')}{E_{k',\ell}} \tanh{\left( \frac{E_{k',\ell}}{2k_B T}\right)}
    \label{eq:self_eta_l}
\end{align}
with
\begin{align}
    E_{k,\ell} = \sqrt{\left( \varepsilon_{k} - \mu \right)^2 + \eta^2_\ell(k)}
\end{align}

Around $T_c$, $~\eta_\ell(k, T_c) \ll \left( \varepsilon_{k} - \mu \right)$, and we can simplify Eq.(\ref{eq:self_eta_l}) as
\begin{align}
    \eta_\ell(k, T_c) = - \int_0^\infty \frac{k' dk'}{4 \pi} \tilde{V_\ell}(k, k') \frac{\tanh{\left( \frac{\left| \varepsilon_{k'} - \mu \right|}{2k_B T_{c,\ell}}\right)}}{\left| \varepsilon_{k'} - \mu \right|} \eta_\ell(k', T_c)
    \label{eq:self_eta_Tc}
\end{align}
The above equation has two unknowns $T_{c,\ell}$ and $\eta_\ell(k', T_c)~ \forall k'$; thus we need additional constraint. We can derive a scalar equation from Eq.(\ref{eq:self_eta_Tc}) by multiplying $\int_0^\infty dk~ \eta_\ell(k, T_c)$ on both sides and we get a resultant equation
\begin{align}
    1 + \frac{\int_0^\infty dk \int_0^\infty \frac{k' dk'}{4 \pi}~ \eta^{(n)}_\ell(k)~ \tilde{V_\ell}(k, k') \frac{\tanh{\left( \frac{\left| \varepsilon_{k'} - \mu \right|}{2k_B T^{(n)}_{c,\ell}}\right)}}{\left| \varepsilon_{k'} - \mu \right|}~ \eta^{(n)}_\ell(k')}{\int_0^\infty dk \left| \eta^{(n)}_\ell(k) \right|^2} = 0
    \label{eq:Tc_step_n}
\end{align}
In writing the above equation, for notational brevity, we omit $T_c$ in the functional form of SC order $\eta_\ell(k, T_c) \equiv \eta_\ell(k)$.

We solve the self-consistent equations \ref{eq:self_eta_Tc} and \ref{eq:Tc_step_n} in two steps. First, we find $T^{(n)}_{c,\ell}$ using Eq.(\ref{eq:Tc_step_n}) at each iteration step $n$, with known $\eta^{(n)}_\ell(k)$ values. Next, using $T^{(n)}_{c,\ell}$ and $\eta^{(n)}_\ell(k)$, we find  $\eta^{(n+1)}_\ell(k)$ from equation \ref{eq:self_eta_Tc} as
\begin{align}
    \eta^{(n+1)}_\ell(k) = - \int_0^\infty \frac{k' dk'}{4 \pi} \tilde{V_\ell}(k, k') \frac{\tanh{\left( \frac{\left| \varepsilon_{k'} - \mu \right|}{2k_B T^{(n)}_{c,\ell}}\right)}}{\left| \varepsilon_{k'} - \mu \right|} \eta^{(n)}_\ell(k')
    \label{eq:eta_step_n}
\end{align}
We repeat this process until we get a converged $\eta_\ell(k)~ \forall k$ and report $T_{c,\ell}$ corresponding to the self-consistent order parameter. The values of  $\eta^{(0)}_\ell(k)~ \forall k$ are initialized using a random distribution of zero mean and a small standard deviation.

At $T=0$, using $\tanh({\infty}) \rightarrow 1$ in Eq.(\ref{eq:self_eta_l}), the zero temperature SC order parameter is found by solving the self-consistency equation
\begin{align}
    \eta^{(n+1)}_\ell(k,0) = - \int_0^{\infty} \frac{k' dk'}{4 \pi} \frac{\tilde{V}_\ell(k, k') ~\eta^{(n)}_\ell(k',0)}{\sqrt{\left( \varepsilon_{k'}-\mu \right)^2 + \left( \eta^{(n)}_\ell(k',0) \right)^2}}
    \label{eq:self_eta_zero}
\end{align}

\begin{figure}
    \centering
    \begin{subfigure}{0.45\linewidth}
        \centering
        \caption{}
        \includegraphics[width=\linewidth,height=5.5cm]{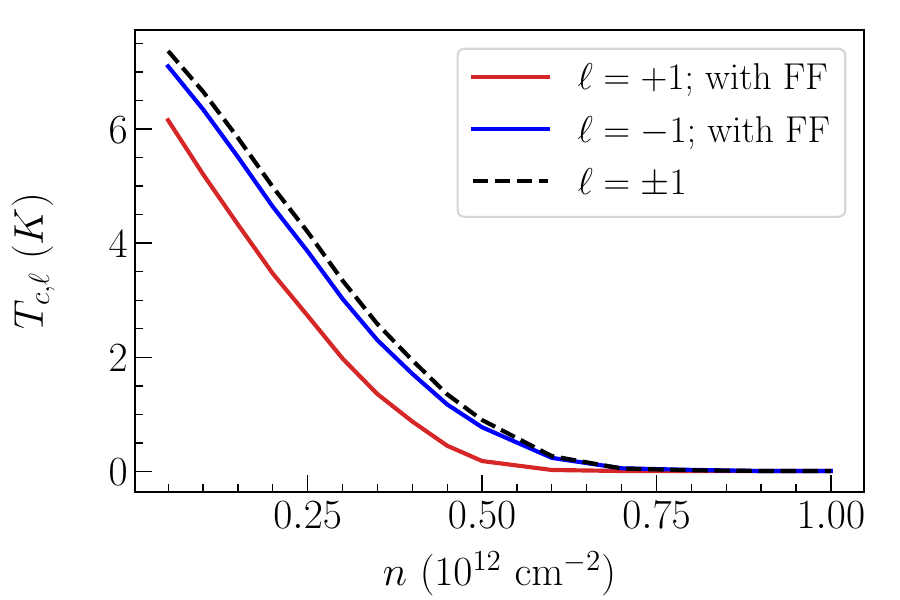}
    \end{subfigure}
    \hspace{0.5cm}
    \begin{subfigure}{0.45\linewidth}
        \centering
        \caption{}
        \includegraphics[width=\linewidth,height=5.5cm]{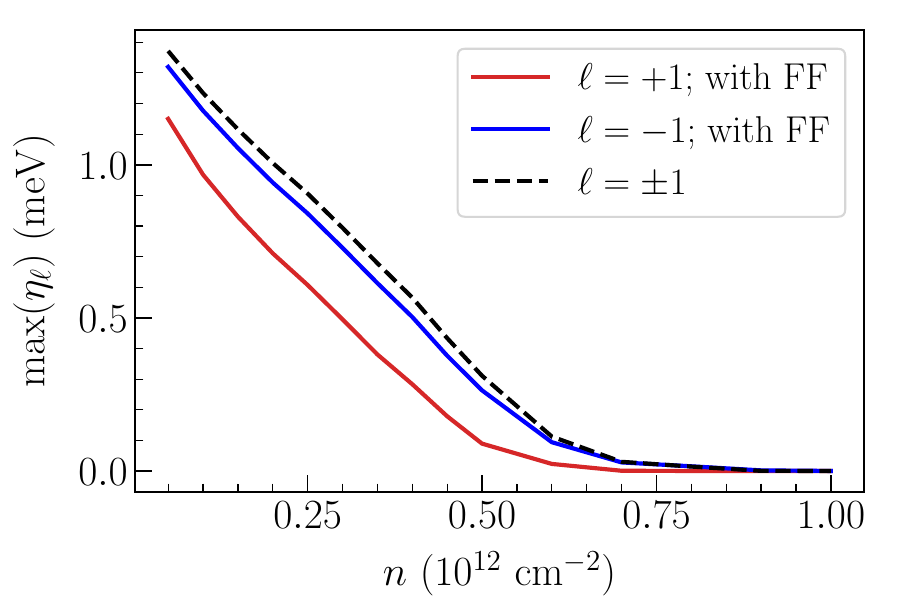}
    \end{subfigure}
    \caption{(a) Superconducting critical temperature $T_{c,\ell}$, (b) maximum of the self-consistent superconducting order $ \eta_\ell(k) $ at $T=0$ for the angular momentum channel $\ell$, with increasing electron density $n$ for metallic gate separation $d_\mathrm{reg}=30$ nm and at the flat-band limit with $E_0=E_0^{\mathrm{flat}}$. Blue and red lines correspond to solutions having interaction with appropriate form factors (FFs).}
    \label{fig:Tc_eta_comparison_with_FF}
\end{figure}

\section{Revised mean-field equation for the interaction with full projection terms}

In the undriven system, we consider an effective two-band model given by the spin-orbit coupling terms for the surface states of a 3D TI. In the presence of driving, these will generate two Floquet bands within the first Floquet Brillouin zone. In this work, we have small electron densities that will populate states near the band edges of the lower Floquet band. Thus, we will consider a single-band problem while keeping the lower band and projecting the interaction term into this pseudospin polarized band.

The superconducting properties will be affected in two stages if we keep the projection terms while assuming the interaction between electrons in the lower Floquet band.

First, the bare charge susceptibility of the electrons is modified by 
\begin{align}
    \chi_c^0(\vq, i \Omega_n) = \frac{1}{A} \sum_{\vk} |\langle u_{\vk} | u_{\vk+\vq} \rangle|^2~ \frac{n_F(\varepsilon_{\vk}) - n_F(\varepsilon_{\vk+\vq})}{\varepsilon_{\vk} - \varepsilon_{\vk+\vq} + i \hbar \Omega_n}
    \label{eq:modified_chi}
\end{align}
where $|u_{\vk} \rangle$ is the single-particle wave function for the lower Floquet band. This will enter while calculating the effective screened interaction within the RPA approximation.

Second, the mean-field equation \ref{eq:MF_SC_eq} is modified by appropriate form factors
\begin{align}
    \Delta(\vk) = - \frac{1}{A} \sum_{\vk'} \tilde{V}(\vk-\vk') \langle u_{\vk}| u_{\vk'} \rangle \langle u_{-\vk}| u_{-\vk'} \rangle \frac{\Delta(\vk')}{2 E_{\vk'}} \tanh \left( \frac{E_{\vk'}}{2k_B T} \right)
    \label{eq:modified_SC_eq}
\end{align}
We can include the form factors in angular decomposition of the effective interaction as
\begin{align}
    \tilde{V}_\ell(k, k') = \int_0^{2\pi} \frac{d\theta}{2\pi}~ e^{i\ell \theta}~ \langle u_{\vk}| u_{\vk'} \rangle \langle u_{-\vk}| u_{-\vk'} \rangle  \tilde{V}(\vk-\vk') ~,~ \cos{\theta}=\frac{\vk.\vk'}{k~ k'}
    \label{eq:modified_angular_decomposition}
\end{align}
With this, the order parameter $\eta_\ell(k)$ for each angular momentum channel takes the same functional form as in Eq.(\ref{eq:self_eta_l}).

The form factors $f( \vk,\vk' )=\langle u_{\vk}| u_{\vk'} \rangle$ are usually complex, so this may break degeneracy between superconducting orders for the angular momentum channels $\pm \ell$ with $\ell \in \mathbb{Z}^+$. As shown in Fig.(\ref{fig:Tc_eta_comparison_with_FF}), within our set-up, we find that while starting from the fully projected interaction by retaining all form factors, the critical temperature $T_c$ as well as the maximum of zero-temperature superconducting order $\eta(k)$ for $\ell=-1$ is elevated over $\ell=1$. The comparison with the resulting $T_c$ and maximal $\eta(k)$ for $\ell=\pm 1$ channels without any form factors in the projected interaction is shown with the black dashed line, suggesting a small deviation. This justifies our assumption of electrons in the lower band to be interacting via pure Coulomb interaction, and the form factors, which inherit the band chirality, only favour a particular chirality for the SC order. We find a maximum difference of $10 \%$ in the $T_c$ for the largest SC order, with and without form factors in the projected interaction term.

\begin{figure}
    \centering
    \begin{subfigure}{0.45\linewidth}
        \centering
        \caption{}
        \includegraphics[width=\linewidth,height=5.5cm]{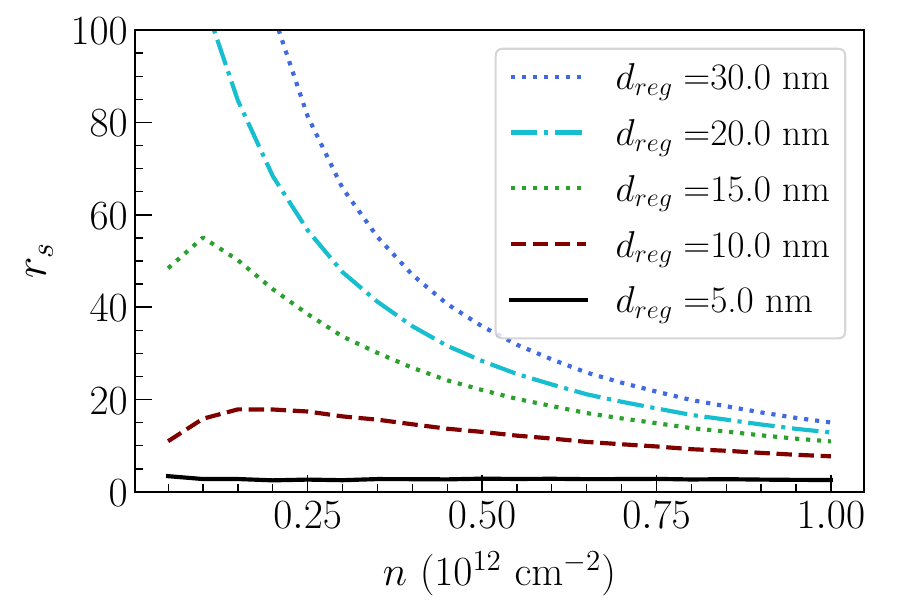}
    \end{subfigure}
    \hspace{0.5cm}
    \begin{subfigure}{0.45\linewidth}
        \centering
        \caption{}
        \includegraphics[width=\linewidth,height=5.5cm]{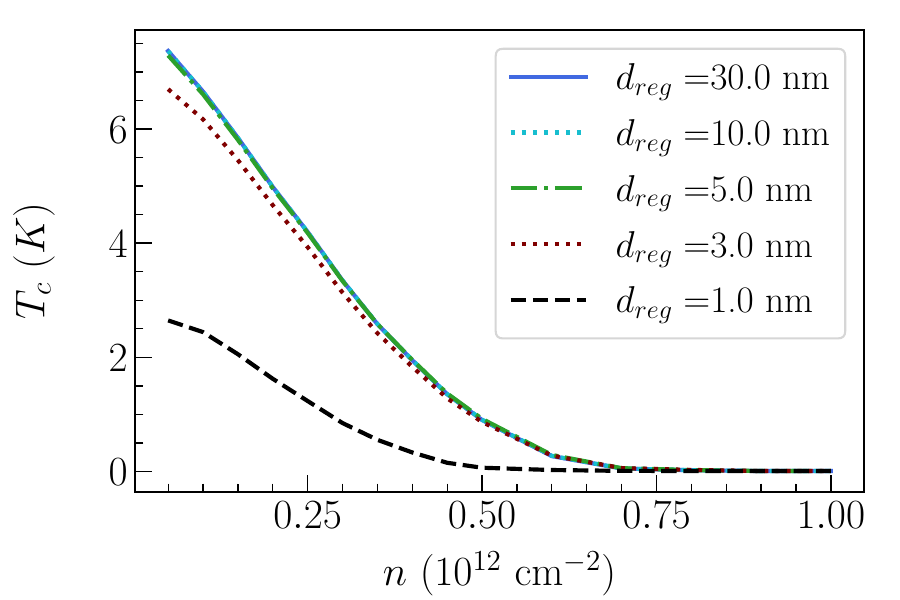}
    \end{subfigure}
    \caption{(a) Evolution of gas parameter $r_s$ and (b) superconducting critical temperature $T_c$ for $l =1$ with metallic gate separation $d_\mathrm{reg}$, from the top surface of 3D TI at the flat-band limit with $E_0=E_0^{\mathrm{flat}}$.}
    \label{fig:rs_Tc_with_dreg}
\end{figure}

\begin{figure}
    \centering
    \includegraphics[width=0.6\linewidth,height=7cm]{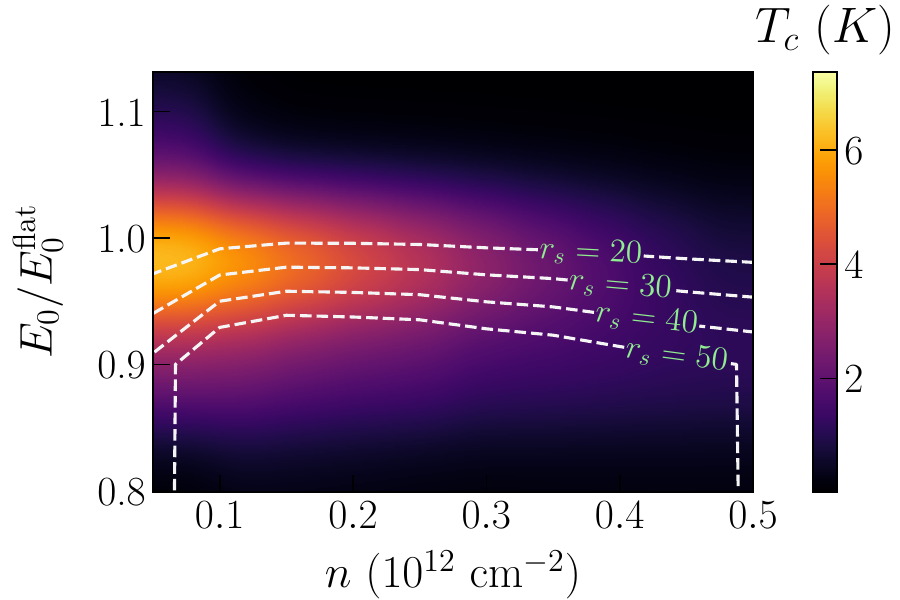}
    \caption{Phase diagram with superconducting critical temperature $T_c$ for $l =1$ as a function of electric field strength $E_0$ and surface electron density $n$ for $d_\mathrm{reg}=10$ nm, with $E_0^{\mathrm{flat}}=\frac{\hbar \omega}{e} \sqrt{\frac{\hbar \omega}{2D}}$. The contour lines are for different gas parameters $r_s$. The region having $r_s \geq 40$ is usually suggestive of a Wigner-crystal phase.}
    \label{fig:Tc_with_rs_contour}
\end{figure}

\section{Effect of the metallic gate separation}

As shown in Fig.(\ref{fig:rs_Tc_with_dreg})(a), we see that the gas parameter $r_s$ is very sensitive to the metallic gate separation $d_\mathrm{reg}$ from the top surface, which is illuminated by circularly polarized light. Usually, when $r_s \geq 40$, the 2D electron gas is susceptible to forming a Wigner-crystal (WC) phase. Here, we find that when $d_\mathrm{reg} \leq 10$ nm and close to the flat-band limit, for the surface electron density $n$ considered, the region has $r_s \leq 20$, thus suggesting a very low probability of forming the WC phase.

Whereas, as shown in Fig.(\ref{fig:rs_Tc_with_dreg})(b), $T_c$ is less sensitive to $d_\mathrm{reg}$, and only when $d_\mathrm{reg} \leq 5$ nm, we see that $T_c$ starts to decrease significantly from $d_\mathrm{reg} \rightarrow \infty$ results for the surface electron densities considered. Thus, having $d_\mathrm{reg} \approx 10$ nm in our set-up will considerably favour the SC phase over the WC phase. 

To highlight how regions of different $r_s$ compare with the $T_c$, we show a phase diagram for $d_\mathrm{reg}=10$ nm, in Fig.(\ref{fig:Tc_with_rs_contour}) by plotting the self-consistent $T_c$ value for the angular momentum channel $\ell=1$, in the parameter space of $E_0$ and $n$. The value of the gas parameter $r_s$ (see main text for the definition) is shown by plotting different contour lines. As we can see, there is a significant region of parameter space with a finite $T_c$ for which $r_s \leq 40$, suggesting the possibility of a superconducting ground state over the competing WC phase.